# Observation of ferromagnetism above 900 K in Cr-GaN and Cr-AlN

H.X. Liu,[a)] Stephen Y. Wu,[a)] R.K. Singh,[a)] Lin Gu,[b)] David J. Smith,[b),c)] N.R. Dilley,[d)]

L. Montes,[d)] M.B. Simmonds,[d)] and N. Newman,[a),e)]

[a)] Department of Chemical and Materials Engineering, Arizona State University, Tempe, Arizona 85287

[b)] Center for Solid State Science, Arizona State University, Tempe, Arizona 85287

[c)] Department of Physics and Astronomy, Arizona State University, Tempe, Arizona 85287

[d)] Quantum Design, San Diego, California 92121

We report ferromagnetism at over 900 K in Cr-GaN and Cr-AlN thin films. The magnetic properties vary as a function of Cr concentration with 60%, and 20%, of the Cr being magnetically active at 3% doping in GaN, and 7% in AlN, respectively. In the GaN sample with the highest magnetically active Cr (60%), channeling Rutherford Backscattering indicates that over 70% of Cr impurities are located on substitutional sites. These results give indisputable evidence that substitutional Cr defects are involved in the magnetic behavior. While Cr-AlN is highly resistive, Cr-GaN exhibits properties characteristic of hopping conduction including $T^{1/2}$ resistivity dependence and small Hall mobility (0.06 cm$^2$/V·s). A large negative magnetoresistance is attributed to the influence of the magnetic field on the quantum interference between the many paths linking two hopping sites. The results strongly suggest that ferromagnetism in Cr-GaN and Cr-AlN can be attributed to the double exchange mechanism as a result of hopping between near-midgap substitutional Cr impurity bands.

[e)] Corresponding author. E-mail address: nathan.newman@asu.edu

Dilute magnetic semiconductors (DMSs) are attractive materials for applications in spin-dependent electronic devices.[1] In these structures, the electron charge and spin are utilized simultaneously to create new functionalities. Such devices offer a unique opportunity to integrate conventional semiconductor and ferromagnetic technologies. The original report of ferromagnetism in Mn-doped GaAs above 100 K aroused intense interest in theoretical and experimental studies of related III-V compound semiconductors doped with magnetic impurities.[2]

For useful practical applications, ferromagnetism must be achieved above room temperature. Recent reports of room temperature ferromagnetism include Mn-GaN,[3] Mn-AlN,[4] Cr-GaN,[5] and Cr-AlN.[6] (Ga,Mn)N films grown by molecular beam epitaxy were reported to be ferromagnetic above room temperature, with a Curie temperature, $T_c$, of 940 K.[7] Another study of the same system synthesized using solid-state diffusion reported a $T_c$ in the range of 220-370 K,[3] while bulk Cr-doped GaN fabricated using the sodium flux method was reported to have a $T_c$ of 280 K.[5]

The search for the physical mechanism responsible for the observed ferromagnetic properties and the question of the applicability of the classical magnetic models have become topics of intense interest. This letter reports the observation of ferromagnetism in Cr-GaN and Cr-AlN above 900 K, and describes the structural, electrical and magnetic properties of the materials.

The Cr-GaN and Cr-AlN films were grown on 6H-SiC (0001) and sapphire (0001) substrates in a reactive molecular beam epitaxy system. Structural properties were characterized using X-ray diffraction (XRD) [Rigaku D/MAX-IIB] and transmission electron microscopy (TEM) [JEOL 4000EX]. The magnetic properties were characterized from 10K to 350K with a SQUID (superconducting quantum interference device) magnetometer (SHE VTS900) and from 300 K to 950 K with a Quantum Design vibrating sample magnetometer (VSM) equipped with the recently developed oven option for the Physical Property Measurement System (PPMS). Magnetic fields were applied parallel to the film plane during susceptibility measurements and perpendicular to the film during magnetoresistance (MR) measurements. The diamagnetic background contributions originating from the substrate and the sample holder were subtracted from the measured magnetic moment to infer the magnetic properties of the deposited thin films. Additional details of the growth and characterization experiments are given elsewhere.[6]

The measured magnetic moments per impurity dopant in Cr-doped AlN vary significantly with Cr concentration range, as shown in Fig. 1. The film with 7% Cr has the highest saturation magnetic moment ($M_s$) of 0.6 $\mu_B$/Cr, which indicates that 20% of the Cr is magnetically active for the AlN sample, when compared to the expected magnetic moment of 3 $\mu_B$/Cr atom.[11] Cr-doped GaN shows a similar trend, although the peak magnetic moment is found at a smaller dopant concentration. For this system, a magnetic moment of 1.8 $\mu_B$/Cr atom is found at a 3% Cr concentration, indicating that 60% of Cr is magnetically active. There is also more variation in the percentage of magnetically active Cr between samples with similar concentrations between different depositions.

The fabrication of some samples with excellent structural quality has enabled the use of advanced channeling ion probe techniques. In a recently synthesized Cr-doped GaN sample, the angular dependence of ion channeling using RBS was measured and a $\chi_{min}$ of 0.03% for Ga and 0.30% for Cr were determined, indicating that over 70% of Cr is on the substitutional sites. Subsequent magnetic measurements indicated that this sample has the highest level of magnetically active Cr (60%) and a Curie temperature over 900 K. These results give indisputable evidence that substitutional Cr is strongly involved in the observed ferromagnetism. This sample was unusual in that it exhibited a decrease in the magnetic moment following the magnetic measurements by over a factor of 2. This drop was consistently observed for measurements of different pieces cut from the same sample.

The lattice constants of Cr-GaN decrease with increasing Cr concentration. The variation in the c-axis lattice parameter follows a nearly linear variation from the undoped value of 5.18053 Å to 5.17198 Å for 7.4% Cr doping. Due to the similarity in the *c*-plane lattice constants of AlN and the SiC substrate, accurate measurement of the lattice parameters of Cr-AlN is difficult. Nevertheless, a similar trend is found, but with significantly more scattering. This variation of lattice constant with doping concentration is characteristic of the introduction of impurity point defects, rather than secondary phases. A similar trend is found in Mn-doped GaN that has been attributed to the incorporation of substitutional impurities on the group-III site.[8] Figure 2(a) shows the field dependence of the magnetization (*M-H*) for GaN doped with 2% Cr at temperatures of 325 K and 800 K. Hysteresis loops are observed at both temperatures, indicating that the film is ferromagnetic. The coercive field is ~100 Oe at 325K and decreases to ~60 Oe at 800 K. Figure 2(b) shows the *M-H* curve of 7% Cr doped AlN. The coercive field is ~120 Oe at 300 K and 70 Oe at 800K. The temperature dependence of magnetization (*M-T*) is shown in Fig. 2(c). Hysteresis in the *M-H* magnetization curves is still observed above 900 K, confirming that both Cr-GaN and Cr-AlN samples have Curie temperatures above 900 K.

To perform electrical measurements, Cr-GaN and Cr-AlN thin films were grown on insulating sapphire substrates.[9] The Cr-AlN sample has a resistivity greater than $10^3$ $\Omega$.cm, whereas the Cr-GaN sample is quite conductive. Fig. 3(a) shows the resistivity of a 4% Cr-doped GaN that exhibits ferromagnetic properties, with 3% Cr magnetically active at room temperature.

The thermally activated process follows the exponential law, $R=R_0\exp[(T_0/T)^{1/2}]$, which is characteristic of variable range hopping between localized states with a Coulomb gap.[10] We attribute the conduction in the GaN films to variable range hopping in the Cr impurity bands. Cr is known experimentally,[11] and predicted theoretically,[12] to form a deep level in the bandgap of GaN at an energy of ~2.0 eV from the valence band. A similar deep level is also predicted for AlN.[13]

The electron carrier density and Hall mobility at 300 K were measured to be $1.4\times10^{20}$ /cm$^3$ and 0.06 cm$^2$/V·s for magnetic fields up to 5 T. The extremely low mobility is related to the observed hopping conductivity. This low value also prevented us from having sufficient measurement accuracy at small magnetic fields to investigate the anomalous Hall effect. Note that the carrier concentration is similar in magnitude to the measured concentration of magnetically-active Cr, $4.9\times10^{19}$/cm$^3$. Since the majority $e$ level of Cr$^{3+}$ is completely filled with two electrons,[11] the remaining electrons that partially fill the $t_2$ level contribute to conduction. Compensating defects in the III-N compounds could increase or decrease the amount of filling of this level. The reason for the quantitative difference is unclear although it may primarily result from an uncertainty in the Hall measurement as a result of localized trapping and magnetic-field-dependent scattering influencing the transport.

Figure 3(b) shows the temperature dependence of the magnetotransport properties of Cr-GaN with a magnetic field up to 7 T applied perpendicular to the sample plane. The sheet resistance showed a strong negative MR from 2K to 300K. The relative change of resistance in magnetic field ($\Delta R/R$) is -3% at 10K, gradually decreasing in magnitude with increasing temperature. The deviation from the parabolic behaviour at high fields and low temperatures is attributed to the conventional path-length-related positive MR, characteristic of traditional semiconductor transport. Negative MR has been observed in other dilute magnetic semiconductors.[12] We attribute the negative MR observed in our films to a mechanism originally proposed by Sivan, Entin-Wohlman, and Imry[14] for non-magnetic semiconductors that takes into account the influence of the magnetic field on the quantum interference between the many paths linking two hopping sites.[15] In the variable range hopping regime and at small magnetic fields, the MR obeys the expression $\Delta R/R = T^{-3/2}B^2$, which is expected in the presence of a Coulomb gap in the density of states,[16] and consistent with the result of temperature dependent resistance. Recently, another transport model[17] has been developed for the case of magnetic semiconductors. This model considers the fluctuations caused by the local exchange interaction between the carrier and magnetization, which introduce additional energy disorder between different carrier hopping sites. The predictions of this model also are found to fit our data.

The Cr-GaN and Cr-AlN materials are ideal for the exploration of the ferromagnetic properties of DMSs since essentially all of the potentially secondary phases, including Cr metal, CrN, $Cr_xGa$ and $Cr_xAl$ alloys, are not ferromagnetic. Only $Cr_2N$ has been reported to be ferromagnetic,[18] although there is very little data on this material and even its Curie temperature has not been reported. Extensive observations by TEM revealed that the Cr-AlN films were epitaxial and single phase. TEM, XRD as well as electron-energy-loss spectroscopy (EELS) do not detect any possible secondary phases in Cr-AlN.[6] In contrast, both XRD and TEM studies indicated that the Cr-GaN films contained very small amounts (<0.2% by volume) of the CrN phase. No evidence for the formation of $Cr_2N$ was found. To address the influence of these compounds on our results, we grew Cr-N films using growth conditions identical to those used for Cr-doped III-N films but without the group-III flux. Characterization of the Cr-N films indicated that the dominant $Cr_2N$ phase comprised ~99.8% of the material, while only ~0.2% was the minority CrN phase. The volume fraction of the CrN phase is fortuitously similar to that in the Cr-doped GaN films. Magnetization measurements of the Cr-N film did not show hysteresis over the temperature range from 10K to 900K and the signals were 1-2 orders lower than those of Cr-doped GaN and AlN. Careful comparisons with measurement background signals suggest that neither $Cr_2N$ nor CrN is ferromagnetic. These findings, combined with the varied magnetic and structural properties of Cr-doped GaN and AlN, allow us to rule out the role of any known secondary phases in contributing to the magnetic properties. We point out that other Cr – Group V compounds such as zincblende CrAs and CrSb,[19] and hexagonal MnAs,[20] have Curie temperatures above room temperature. Nanocrystalline $Cr_xN$ phases have also been theoretically predicted to have a Curie temperature above room temperature,[21] It is difficult to conclusively rule out the presence of a minute amount of extraneous secondary phases using techniques such as XRD and TEM alone. Our use of RBS channelling does, however, prove that the majority of Cr is on substitutional lattice sites. It is not clear if non-substitutional Cr plays a significant role in the ferromagnetic properties.

Currently, there is no consensus on the nature of ferromagnetism in DMS materials. Several theoretical models have been proposed to understand this magnetic ordering. A free-carrier-mediated model has been proposed to explain the magnetic properties of Mn-doped III-V semiconductors;[22] the presence of free holes due to the shallow acceptor nature of $Mn^{2+}+h^+$ in GaAs might be a reasonable approximation, but it most certainly does not apply to the ferromagnetism in Cr- and Mn-doped GaN and AlN since these defects form near midgap deep levels.[11,13] The Cr $d$ level in GaN and AlN is split by exchange and the crystal field into a threefold degenerate $t_2$ and a doubly degenerate $e$ level.[13] The majority $t_2$ would be expected to be $1/3$ filled for Cr and $2/3$ filled for Mn. The partial filling of the majority $t_2$ level suggests a ferromagnetic ground state associated with the double exchange mechanism. In such cases, the Fermi level is pinned within the defect bands. The observed hopping transport characteristics, in conjunction with the similarity between the concentration of transport electrons and the density of magnetically active Cr atoms, support this theory.

In conclusion, high-quality Cr-GaN and Cr-AlN thin films exhibit ferromagnetism with a Curie temperature above 900 K. The magnetic properties are found to strongly vary as a function of magnetic impurity concentration with the best characteristics resulting at 3 % Cr in GaN and 7 % Cr in AlN. Extensive structural characterization using XRD, TEM and ion channeling gives indisputable evidence that substitutional Cr defects are involved in the magnetic behavior.

This work was supported by the Defense Advanced Research Projects Agency (DARPA) and administered by the Office of Naval Research (Contract No. N00014-02-1-0598). The authors thank Raghu Gandikota and Brett Strawbridge for their assistance in performing electrical measurements and Barry Wilkens for his assistance in angular scan. We also thank David Look, Mark van Schilfgaarde, Art Freeman, John Rowell and David Ferry for enlightening discussions.

Figure Captions

FIG. 1 Magnetic properties of Cr-AlN grown at 700 °C and 800 °C with various Cr concentrations. Lines are included to illustrate the general trends observed.

FIG. 2(a) Magnetic field dependence of magnetization of (a) 2% Cr-doped GaN at 325 K and 800 K, (b) 7% Cr-doped AlN at 300 K and 800 K. (c) Temperature dependence of magnetization of 2% Cr-doped GaN and 7% Cr-doped AlN.

FIG. 3. Temperature dependent transport measurements of Cr-GaN deposited on sapphire (a) Resistivity ($\rho$). Insert shows comparison between experimental data and functional relationship expected for variable range hopping. (b) Magnetoresistance ($\Delta R/R$), the relative change of sheet resistance in a magnetic field.

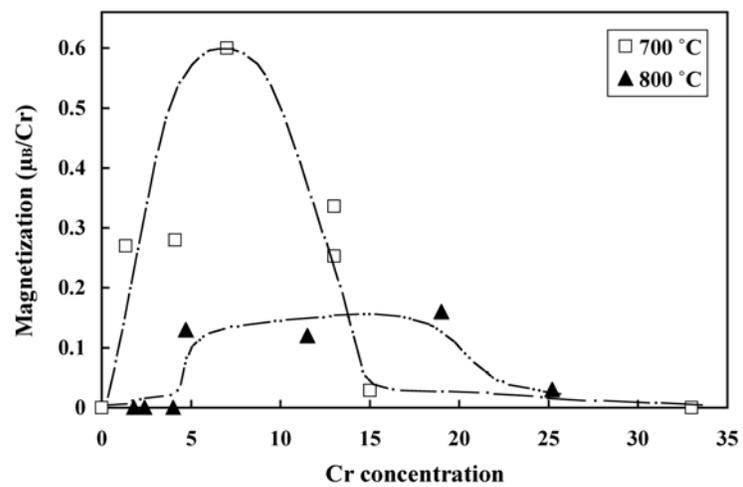

**FIG. 1. – Liu *et al.***

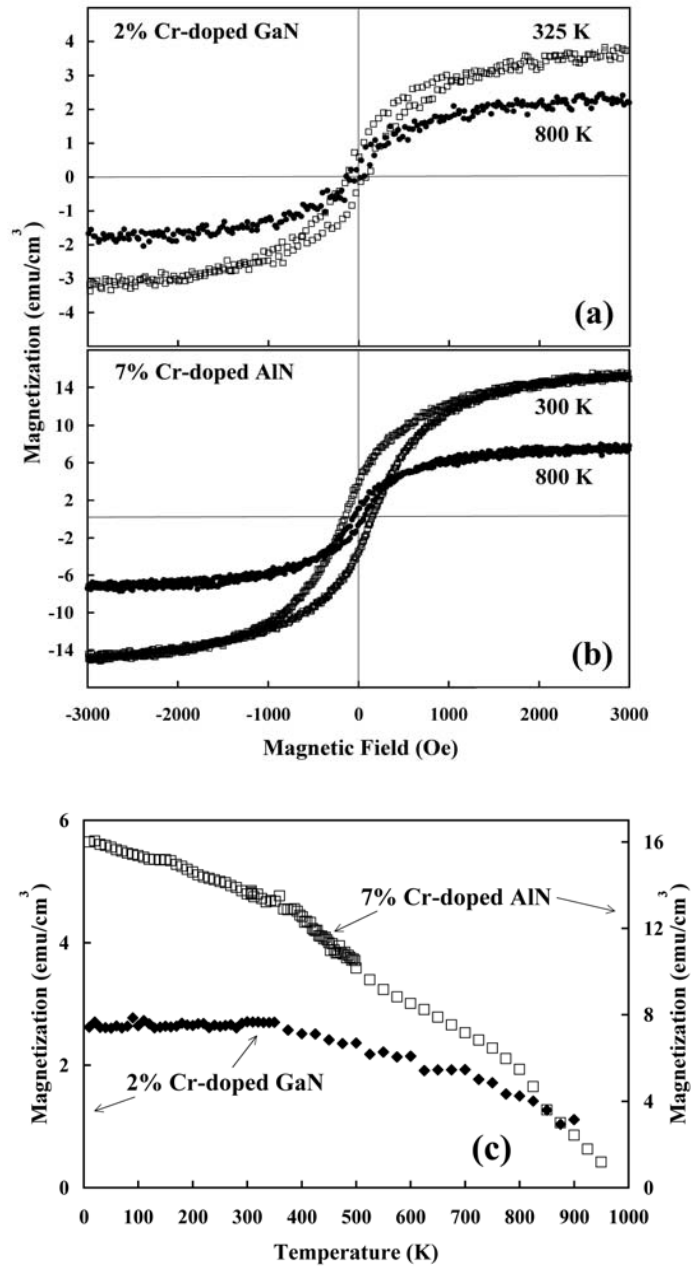

**FIG. 2 – Liu *et al.***

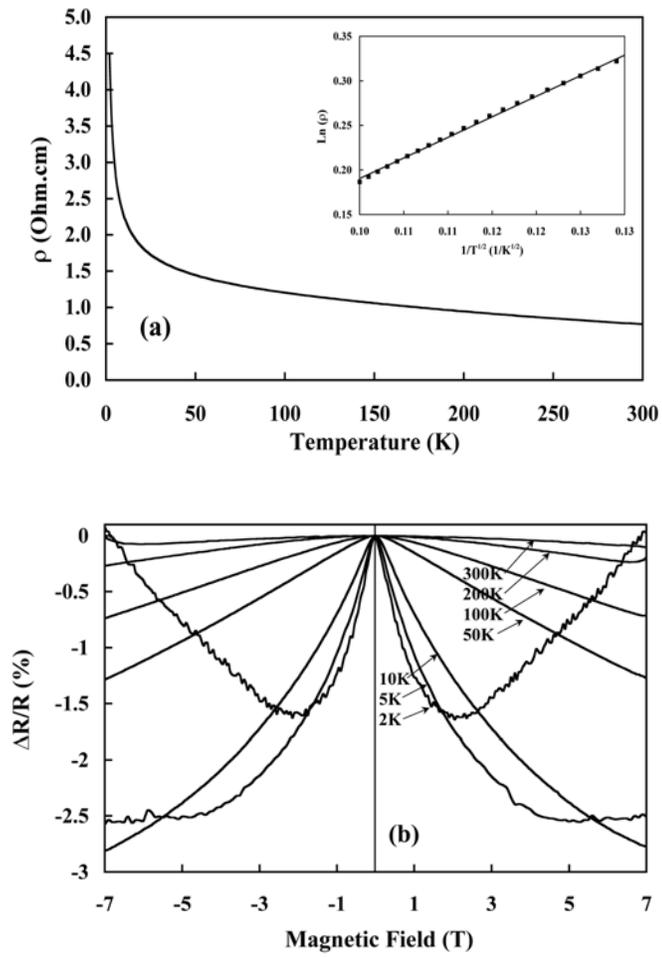

**FIG. 3. – Liu *et al.***